\documentclass[twocolumn,nopacs,floatfix,amsmath,nofootinbib,amssymb,preprintnumbers,floatfix]{revtex4}
\usepackage{longtable,lscape,booktabs}
\usepackage{txfonts}
\usepackage{overpic}
\usepackage{amssymb}
\usepackage{indentfirst}
\usepackage{feynmf}
\usepackage{slashed}
\usepackage{cases}
\usepackage{color}
\usepackage{multirow}
\usepackage{ulem}
\usepackage{subfigure}
\usepackage{graphicx,color,dcolumn,bm}
\usepackage[colorlinks,
            citecolor=blue,
            anchorcolor=red,
            menucolor=red,
            linkcolor=red,
            filecolor=red,
            runcolor=red,
            urlcolor=blue,
            frenchlinks=red]{hyperref}

\begin{document}

\title{Study on the structure of the $Z_{c}(3900)$ state}

\author{Yan Ma$^{1,2,3}$}
% \email{mycurry302022@163.com}
\author{De-Shun Zhang$^2$}
\author{Cheng-Qun Pang$^{1,3,4}$}
\email
 [corresponding author: ]
{pcq@qhnu.edu.cn}
\author{Zhi-Feng Sun$^{2,3,5}$}
\email
 [corresponding author: ]
{sunzf@lzu.edu.cn}

\affiliation{$^1$College of Physics and Electronic Information Engineering, Qinghai Normal University, Xining 810000, China
\\$^2$School of Physical Science and Technology, Lanzhou University, Lanzhou 730000, China
\\$^3$Lanzhou Center for Theoretical Physics, Key Laboratory of Theoretical Physics of Gansu Province, and Key Laboratory of Quantum Theory and Applications of the Ministry of Education, Lanzhou University, Lanzhou, 730000, China
\\$^4$Joint Research Center for Physics,Lanzhou University and Qinghai Normal University,
Xining 810000, China
\\$^5$Frontiers Science Center for Rare Isotopes, Lanzhou University, Lanzhou, Gansu 730000, China}

\date{\today}
\begin{abstract}
In this work, we studied the $Z_{c}(3900)$ state within the framework of effective field theory. We firstly show the construction of the Lagrangian describing meson-meson-meson and meson-diquark-diquark interactions. By using the Feynman rule, we calculate the effective potentials corresponding to the coupled channels of $D\bar{D}^{*}/D^{*}\bar{D}$ and $S_{cq}\bar{A}_{cq}/A_{cq}\bar{S}_{cq}$ with $S_{cq}$ ($A_{cq}$) the scalar (axial vector) diquark composed of $c$ and $q$ quarks. After solving the Bethe-Salpeter equation of the on-shell parametrized form and compare our numerical results with the experimental mass and width of $Z_{c}(3900)$, we find that the $Z_{c}(3900)$ state can be explained as the mixture of $D\bar{D}^{*}/D^{*}\bar{D}$ and $S_{cq}\bar{A}_{cq}/A_{cq}\bar{S}_{cq}$ components. 
\end{abstract}

\maketitle

\section{Introduction} 
Since the discovery of new states that do not fit in the conventional definition of hadron as $q\bar{q}$ meson and $qqq$ baryons in the early 2000's, hadronic physics has definitively entered a new era. For review, see Refs. \cite{Chen:2016qju,Liu:2019zoy,Hosaka:2016pey,Lebed:2016hpi,Esposito:2016noz,Guo:2017jvc,Albuquerque:2018jkn,Brambilla:2019esw,Guo:2019twa,Meng:2022ozq,Bicudo:2022cqi,Liu:2024uxn,Chen:2022asf,Dong:2021bvy,Dong:2017gaw,Yang:2020atz}. In this work, we only focus on the $Z_c(3900)$ state. In 2013, BESIII Collaboration observed a new charged state $Z^{+}_{c}(3900)$ for the first time during the $e^{+}e^{-}\longrightarrow \pi ^{+}\pi ^{-}J/\psi $ reaction \cite{BESIII:2013ris}. Soon after that, the Belle Collaboration confirmed the existence of this particle \cite{Belle:2013yex}. In 2014, BESIII Collaboration also observed charged $Z^{+}_{c}(3900)$ on the $D\bar{D}^{*}$ invariant mass spectrum of the charm process $e^{+}e^{-}\longrightarrow \pi ^{\pm}(D\bar{D}^{*})$ \cite{BESIII:2013qmu}. In 2015, BESIII Collaboration observed neutral $Z^{0}_{c}(3900)$ particles in both $e^{+}e^{-}\longrightarrow \pi ^{0}\pi ^{0}J/\psi$ and $e^{+}e^{-}\longrightarrow \pi ^{0}(D\bar{D}^{*})^{0}$ processes \cite{BESIII:2015cld,BESIII:2015ntl}. In 2017, BESIII determined the spin-parity of $Z_{c}(3900)$ as $1^{+}$ \cite{BESIII:2017bua}. The most recently measured mass and width of the $Z_{c}(3900)$ state are given by \cite{BESIII:2020oph}
\begin{eqnarray}
    m&=&3893.1\pm 2.2\pm 3.0 \thinspace\text{MeV}, \nonumber\\
    \Gamma&=&44.4\pm 5.2\pm 14.0\ \text{MeV},\nonumber
\end{eqnarray}
respectively. For other experimental studies of $Z_c(3900)$, one could refer to \cite{D0:2018wyb,BESIII:2019rek,D0:2019zpb}. In Ref. \cite{Chen:2023def}, the authors performed a unified description of the experimental data of the $\pi^+\pi^-$ and $J/\psi\pi^\pm$ invariant mass spectra for $e^+e^-\to J/\psi \pi^+\pi^-$ and the $D^0D^{*-}$ mass spectrum for $e^+e^-\to D^0D^{*-}\pi^+$ at two energy points $E=4.23$ GeV and $4.26$ GeV, in which the pole mass and width of $Z_c(3900)$ are precisely determined as 
\begin{eqnarray}
m&=&3880.7\pm 1.7\pm 22.4 \text{MeV}, \nonumber \\    
\Gamma&=&35.9\pm 1.4\pm 15.3 \text{MeV}.
\end{eqnarray}

After the discovery of $Z_c(3900)$ in 2013, there were many debates on the nature of it. Since the given isospin of $Z_c(3900)$ is 1, it can not be of charmonium structure. On the other hand, due to its mass close to the $D\bar{D}^*/D^*\bar{D}$ threshold, some works interpreted this particle as a molecular state. However, other explanations can not be ruled out, such as tetraquark, kinematical effects and so on. There are a large numbers of papers discussing these topics, here we refer to the reviews \cite{Chen:2016qju,Liu:2019zoy,Olsen:2014qna,Hosaka:2016pey,Lebed:2016hpi,Esposito:2016noz,Guo:2017jvc,Ali:2017jda,Karliner:2017qhf,Albuquerque:2018jkn,Brambilla:2019esw,Guo:2019twa,Meng:2022ozq,Liu:2024uxn,Chen:2022asf,Dong:2017gaw}.

The topics of molecular state have been widely discussed \cite{Chen:2016qju,Chen:2022asf,Dong:2021bvy,Meng:2022ozq,Liu:2024uxn} after the discovery of $X(3872)$ \cite{Belle:2003nnu}, which was proposed based on the study of deuteron. The existence of the tetraquark made of diquark-antidiquark pair (also named diquark-antidiquark state) was suggested in Ref. \cite{Lichtenberg:1967zz} very soon after Gell-Mann proposed the quark and mentioned the possibility of diquarks \cite{Gell-Mann:1964ewy}. Later, many groups studied on this subject, see the reviews \cite{Anselmino:1992vg,Brambilla:2019esw}.

In the present work, we will take into account the mixture of the above two types of exotic hadrons configuration, i.e., the molecular state and diquark-antidiquark state, within the framework of the effective field theory, and try to understand the nature of $Z_c(3900)$. The theoretical method utilized in the present work has been established in Refs. \cite{Cao:2022rjp,He:2024aej,Zhang:2024zbo}, in which $Z_{cs}(4000)$, $Z_{cs}(4220)$, $Z_b(10610)$, $Z_b(10650)$ and $X(4500)$ were well explained as the mixtures of molecular and diquark-antidiquark components. 

The paper is organized as follows. After the introduction, we show the construction of the Lagrangians in Sec. {II}. Then we give the effective potentials and discuss the Bethe-Salpeter equation of the on-shell factorized form in Sec. III. The numerical results and discussion are shown in Sec. {V}. Finally, a brief summary is given in Sec. {VI}.

\section{The construction of the Lagrangians}
In the present work, in order to construct the Lagrangians describing the interactions of charmed meson-charmed meson-light meson, charmed meson-charmed diquark-light diquark and charmed diquark-charned diquark-light meson, the hidden local gauge symmetry (HLS) is taken into account. We take the model following Ref. \cite{Bando:1987br} possessing the $G_{global}\otimes H_{local}$ symmetry, where $G=U(3)_L\otimes U(3)_R$ is the global chiral symmetry and $H=U(3)_V$ is the HLS. In this symmetry, the vector mesons are introduced as the gauge bosons, i.e.,
\begin{eqnarray}
    V_\mu &=&\frac{g_{V}}{\sqrt{2}}\left(
\begin{array}{ccc}
\frac{1}{\sqrt{2}}(\rho^{0}+\omega)&\rho^+&K^{*+}\\
\rho^-&-\frac{1}{\sqrt{2}}(\rho^{0}-\omega)&K^{*0}\\
K^{*-}&\bar{K}^{*0}&\phi
\end{array}
\right)_\mu.
\end{eqnarray}
The basic quantities $\xi_L$ and $\xi_R$ are introduced by dividing $U$ in the chiral perturbation theory as $U=\xi_L^\dag\xi_R$ with $U=e^{i\Phi/f_\pi}$. The transformation properties of the two varibles, i.e., $\xi_L$ and $\xi_R$ under the $G_{global}\otimes H_{local}$ symmetry are
\begin{eqnarray}
    \xi_{L,R}(x)\to \xi_{L,R}^\prime=h(x)\xi_{L,R}g^\dag_{L,R},
\end{eqnarray}
where 
\begin{eqnarray}
    h(x)\in H_{local},\ g_{L,R}\in G_{global}.
\end{eqnarray}
These varibles can be parameterized in the form of
\begin{eqnarray}
    \xi_{L,R}=e^{i\sigma/f_\sigma}e^{\mp i\Phi/(2f_\pi)}.\label{eq4}
\end{eqnarray}
$\Phi$ denotes the Nambu-Goldstone bosons associated with the spontaneous breaking of the global chiral symmetry $G$ 
\begin{eqnarray}
    \Phi=\sqrt{2}\left(
\begin{array}{ccc}
\frac{\sqrt{3}\pi^0+\eta+\sqrt{2}\eta'}{\sqrt{6}}&\pi^+&K^+\\
\pi^-&\frac{-\sqrt{3}\pi^0+\eta+\sqrt{2}\eta'}{\sqrt{6}}&K^0\\
K^-&\bar{K}^0&\frac{-2\eta+\sqrt{2}\eta'}{\sqrt{6}}
\end{array}
\right),
\end{eqnarray}
and $\sigma$ denotes the Nambu-Goldstone bosons which can be absorbed into the gauge bosons.

Two Maurer-Cartan 1-forms are defined as
\begin{eqnarray}
    \alpha_\bot^\mu&=&(\partial_{\mu}\xi_{R}\xi_{R}^{\dag}-\partial_{\mu}\xi_{L}\xi_{L}^{\dag})/2i,\\
    \alpha_\parallel^\mu&=&(\partial_{\mu}\xi_{R}\xi_{R}^{\dag}+\partial_{\mu}\xi_{L}\xi_{L}^{\dag})/2i,
\end{eqnarray}
which transform under $G_{global}\otimes H_{local}$ symmetry as
\begin{eqnarray}
    \alpha_{\bot\mu}&\to&h(x)\alpha_{\bot\mu}h(x)^\dag,\\
    \alpha_{\|\mu}&\to&h(x)\alpha_{\|\mu}h(x)^\dag-i\partial_\mu h(x)h(x)^\dag.
\end{eqnarray}
The covariant derivatives of $\xi_L$ and $\xi_R$ read
\begin{eqnarray}
    D_\mu\xi_{L,R}=\partial_\mu\xi_{L,R}-iV_\mu \xi_{L,R}
\end{eqnarray}
with the vector meson field $V_\mu$ transforms as 
\begin{eqnarray}
    V_\mu\to h(x)V_\mu h(x)^\dag-i\partial_\mu h(x)h(x)^\dag,
\end{eqnarray}
and $D_\mu \xi_{L,R}$ as
\begin{eqnarray}
    D_\mu \xi_{L,R}\to h(x)D_\mu \xi_{L,R}h(x)^\dag.
\end{eqnarray}
Then the covariant 1-form are given by
\begin{eqnarray}
    \hat{\alpha}_{\|\mu}&=&(D_{\mu}\xi_{R}\xi_{R}^{\dag}+D_{\mu}\xi_{L}\xi_{L}^{\dag})/2i,\\
    \hat{\alpha}_{\bot\mu}&=&(D_{\mu}\xi_{R}\xi_{R}^{\dag}-D_{\mu}\xi_{L}\xi_{L}^{\dag})/2i,
\end{eqnarray}
which have the following relations to $\alpha_{\bot\mu}$ and $\alpha_{\|\mu}$
\begin{eqnarray}
     \hat{\alpha}_{\|\mu}&=&\alpha_{\|\mu}-V_{\mu},\\
    \hat{\alpha}_{\bot\mu}&=&\alpha_{\bot\mu}.
\end{eqnarray}
The covariantized 1-forms $\hat{\alpha}_{\|\mu}$ and $\hat{\alpha}_{\bot\mu}$ now transform homogeneously
\begin{eqnarray}
    \hat{\alpha}_{\bot,\|}^\mu\to h(x)\hat{\alpha}_{\bot,\|}^\mu h(x)^\dag.
\end{eqnarray}

Since the considered vertexes involve charmed mesons, we need to deal with the corresponding fields, which are shown below
\begin{eqnarray}
    P&=&(D^0,D^+,D_s^+),\\
    P^*_{\mu}&=&(D^{*0},D^{*+},D_s^{*+})_{\mu}.
\end{eqnarray}
The transformation properties of these fields are chosen as
\begin{eqnarray}
    P\to Ph(x)^\dag, \ P^*_\mu\to P^*_\mu h(x)^\dag.
\end{eqnarray}
The covariant derivatives of $P$ and $P^*_\mu$ are defined as 
\begin{eqnarray}
    D_\mu P&=&\partial_\mu P+iP\alpha_{\|\mu}^\dag=\partial_\mu P+iP\alpha_{\|\mu},\\
    D_\mu P^*_\mu&=&\partial_\mu P^*_\mu+iP^*_\mu\alpha_{\|\mu}^\dag=\partial_\mu P^*_\mu+iP^*_\mu\alpha_{\|\mu},
\end{eqnarray}
which transform as 
\begin{eqnarray}
    D_\mu P&\to&D_\mu Ph(x)^\dag,\\
    D_\mu P^*_\nu&\to& D_\mu P^*_\nu h(x)^\dag.
\end{eqnarray}

On the other hand, the charmed diquark and light diquark fields also need to be considered, which is shown below
\begin{eqnarray}
    S_{c}^{a}=(S_{cu},S_{cd},S_{cs})^{a},\\
    A_{c\mu}^{a}=(A_{cu},A_{cd},A_{cs})_{\mu}^{a},
\end{eqnarray}
\begin{eqnarray}
S^{a}&=&\left(
\begin{array}{ccc}
	0&S_{ud}&S_{us}\\
	-S_{ud}&0&S_{ds}\\
	-S_{us}&-S_{ds}&0
\end{array}
\right)^{a},\\
A_{\mu}^{a}&=&\left(
\begin{array}{ccc}
	A_{uu}&\frac{1}{\sqrt{2}}A_{ud}&\frac{1}{\sqrt{2}}A_{us}\\
	\frac{1}{\sqrt{2}}A_{ud}&A_{dd}&\frac{1}{\sqrt{2}}A_{ds}\\
	\frac{1}{\sqrt{2}}A_{us}&\frac{1}{\sqrt{2}}A_{ds}&A_{ss}
\end{array}
\right)^{a}_{\mu},
\end{eqnarray}
with the superscript $a=1,2,3$ the color index. Note that only the color antitriplets are taken into account, since  the interaction between the quarks of an antitriplet was attractive, while that of a color sextet was repulsive. The transform properties of the diquark fields are listed in the following
\begin{eqnarray}
    S_c^a&\to& S_c^ah(x)^T,\\
    A_{c\mu}^a&\to& A_{c\mu}^ah(x)^T,\\
    S^a&\to& h(x)S^ah(x)^T,\\
    A_{\mu}^a&\to& h(x)A_{\mu}^ah(x)^T.
\end{eqnarray}
The corresponding covariant derivatives are defined as 
\begin{eqnarray}
&D_{\mu}S_{c}^{a}&=\partial_{\mu}S_{c}^{a}-iS_{c}^{a}\alpha_{\|\mu}^{T},\\
&D_{\mu}A^{a}_{c\nu}&=\partial_{\mu}A^{a}_{c\nu}-iA^{a}_{c\nu}\alpha_{\|\mu}^{T},\\
&D_{\mu}S^{a}&=\partial_{\mu}S^{a}-iV_{\mu}S^{a}-iS^{a}V_{\mu}^{T},\\
&D_{\mu}A^{a}_{\nu}&=\partial_{\mu}A^{a}_{\nu}-iV_{\mu}A^{a}_{\nu}-iA^{a}_{\nu}V_{\mu}^{T}.
\end{eqnarray}

The transformation properties of all the quantities discussed above under parity and charge conjugation are listed in TABLE \ref{tab1}.
\begin{table*}
    \caption{The transformation properties of some quantities under parity and charge conjugation denoting by $\mathcal{P}$ and $\mathcal{C}$, respectively.}
    \begin{tabular}{cccccccccccccccccccc}\toprule[1pt]
         & $P$&$P_\mu^*$&$D_\mu P$&$D_\mu P_\nu^*$&$\hat{\alpha}_{\bot\mu}$&$\hat{\alpha}_{\|\mu}$&$V_\mu$&\\\hline
         $\mathcal{P}$&$-P$ &$P^{*\mu}$&$-D^\mu P$&$D^\mu P^{*\nu}$&$-\hat{\alpha}_\bot^\mu$&$\hat{\alpha}_\|^\mu$&$V^\mu$&\\
         $\mathcal{C}$&$P^{\dag T}$ &$-P^{*\dag T}_\mu$&$D_\mu P^{\dag T}$&$-D_\mu P^{*\dag T}_\nu$&$\hat{\alpha}_{\bot\mu}^T$&$-\hat{\alpha}_{\|\mu}^T$&$-V_\mu^T$&\\         
         \hline\hline
         &$S^a$&$A_\mu^a$&$D_\mu S^a$&$D_\mu A_\nu^a$&$S_c^a$&$A_{c\mu}^a$&$D_\mu S_c^a$&$D_\mu A_{c\mu}^a$\\\hline
         $\mathcal{P}$&$S^a$ &$-A^{a\mu}$&$D^\mu S^a$&$-D^\mu A^{a\nu}$&$S_c^a$&$-A_c^{a\mu}$&$D^\mu S_c^a$&$-D^\mu A_c^{a\nu}$\\
         $\mathcal{C}$&$-S^{a\dag T}$ &$A_{\mu}^{a\dag T}$&$-D_\mu S^{a\dag T}$&$D_\mu A_\nu^{a\dag T}$&$-S_c^{a\dag T}$&$A_{c\mu}^{a\dag T}$&$-D\mu S_c^{a\dag T}$&$D_\mu A_{c\nu}^{a\dag T}$\\\bottomrule[1pt]
    \end{tabular}
    \label{tab1}
\end{table*}

With the quantities above and their transformation properties, now we are ready to construct the Lagrangian. The vertices of charmed meson and light meson are given by
\begin{eqnarray}
\mathcal{L}_{1}&=&a_1(iP\hat{\alpha}_{\|\mu}D^{\mu}P^{\dag}+h.c.)+a_2(iP\hat{\alpha}_{\perp\mu}P^{*\mu\dag}\nonumber\\
&&+h.c.)+a_3(\xi^{\mu\nu\alpha\beta}P^{*}_{\nu}\hat{\alpha}_{\perp\alpha}D_{\mu}P^{*\dag}_{\beta}+h.c.)\nonumber\\
&&+a_4(iP^{*}_{\nu}\hat{\alpha}_{\|}^{\mu}D_{\mu}P^{*\nu\dag}+h.c.).\label{eq37}
\end{eqnarray}
The interactions of charmed meson, charmed diquark and light diquark are described by
\begin{eqnarray}
\mathcal{L}_{2}&=&e_{1}(iPD_{\mu}S^aA_{c}^{a\mu\dag}-iA_{c}^{a\mu}D_{\mu}S^{a\dag}P^{\dag})\nonumber\\
&&+e_{2}(iPA^a_{\mu}D^{\mu}S_{c}^{a\dag}-iD^{\mu}S^a_{c}A_{\mu}^{a\dag}P^{\dag})\nonumber\\
&&+e_{3}(\epsilon^{\mu\nu\alpha\beta}PA^a_{\mu\nu}A_{c\alpha\beta}^{a\dag}+\epsilon^{\mu\nu\alpha\beta}A^a_{c\alpha\beta}A_{\mu\nu}^{a\dag}P^{\dag})\nonumber\\
&&+e_{4}(iP_{\mu}^{*}D^{\mu}S^aS_{c}^{a\dag}-iS^a_{c}D^{\mu}S^{a\dag}P_{\mu}^{*\dag})\nonumber\\
&&+e_{5}(\epsilon^{\mu\nu\alpha\beta}P_{\mu}^{*}D_{\nu}S^aA_{c\alpha\beta}^{a\dag}+\epsilon^{\mu\nu\alpha\beta}A^a_{c\alpha\beta}D_{\nu}S^{a\dag}P_{\mu}^{*\dag})\nonumber\\
&&+e_{6}(\epsilon^{\mu\nu\alpha\beta}P_{\mu}^{*}A^a_{\nu\alpha}D_{\beta}S_{c}^{a\dag}+\epsilon^{\mu\nu\alpha\beta}D_{\beta}S^a_{c}A_{\nu\alpha}^{a\dag}P_{\mu}^{*\dag})\nonumber\\
&&+e_{7}(iP_{\mu}^{*}A^{a\mu\nu}A_{c\nu}^{a\dag}-iA^a_{c\nu}A^{a\mu\nu\dag}P_{\mu}^{*\dag})\nonumber\\
&&+e_{8}(iP_{\mu}^{*}A^a_{\nu}A_{c}^{a\mu\nu\dag}-iA_{c}^{a\mu\nu}A_{\nu}^{a\dag}P_{\mu}^{*\dag})\nonumber\\
&&+e_{9}(iP_{\mu\nu}^{*}A^{a\mu}A_{c}^{a\nu\dag}-iA_{c}^{a\nu}A^{a\mu\dag}P_{\mu\nu}^{*\dag}).
\label{eq38}
\end{eqnarray}
The vertices involving two charmed diquarks and a light meson read
\begin{eqnarray}
\mathcal{L}_{3}&=&h_{1}(iS^a_{c}\hat{\alpha}_{\|}^{\mu T}D_{\mu}S_{c}^{a\dag}-iD_{\mu}S^a_{c}\hat{\alpha}_{\|}^{\mu T}S_{c}^{a\dag})\nonumber\\
&&+h_{2}(\epsilon^{\mu\nu\alpha\beta}A^a_{c\mu\nu}\hat{\alpha}_{\|\alpha}^{T}D_{\beta}S_{c}^{a\dag}+\epsilon^{\mu\nu\alpha\beta}D_{\beta}S^a_{c}\hat{\alpha}_{\|\alpha}^{T}A_{c\mu\nu}^{a\dag})\nonumber\\
&&+h_{3}(iA^a_{c\mu}\hat{\alpha}_{\bot}^{\mu T}S_{c}^{a\dag}-iS^a_{c}\hat{\alpha}_{\bot}^{\mu T}A_{c\mu}^{a\dag})\nonumber\\
&&+h_{4}(iA^a_{c\mu}\hat{\alpha}_{\|\nu}^{T}A_{c}^{a\mu\nu\dag}-iA_{c}^{a\mu\nu}\hat{\alpha}_{\|\nu}^{T}A_{c\mu}^{a\dag})\nonumber\\
&&+h_{5}(\epsilon^{\mu\nu\alpha\beta}A^a_{c\mu}\hat{\alpha}_{\bot\nu}^{T}A_{c\alpha\beta}^{a\dag}+\epsilon^{\mu\nu\alpha\beta}A^a_{c\alpha\beta}\hat{\alpha}_{\bot\nu}^{T}A_{c\mu}^{a\dag}).
\label{eq39}
\end{eqnarray}
Note that in Eqs. \eqref{eq38} and \eqref{eq39}, the field strength tensor for the vector mesons and the axial vector diquarks are defined as
\begin{eqnarray}
P^*_{\mu\nu}&=&D_\mu P^*_\nu-D_\nu P^*_\mu,\label{eq40}\\
A_{c\mu\nu}^{a}&=&D_{\mu}A_{c\nu}^{a}-D_{\nu}A_{c\mu}^{a},\label{eq41}\\
A_{\mu\nu}^{a}&=&D_{\mu}A_{\nu}^{a}-D_{\nu}A_{\mu}^{a}.\label{eq42}
\end{eqnarray}
In Eqs. \eqref{eq38} and \eqref{eq39}, the repeated superscripts ``$a$" (the color indices) mean the summation over them. 

\section{THE EFFECTIVE POTENTIALS}
\subsection{The flavor wave functions}
In the present work, what we investigate is the $Z_c(3900)$ state. Henceforth, the quantum number of the considered system is $I^G(J^{PC})=1^+(1^{+-})$. Corespondingly, the coupled channels are $D\bar{D}^*/D^*\bar{D}$ and $S_{cq}\bar{A}_{cq}/A_{cq}\bar{S}_{cq}$. And the wave functions read
\begin{eqnarray}
|{X_{D\bar{D}^{*}/D^*\bar{D}}}\rangle&=&\frac{1}{2}[(|\bar{D}^{*0}D^{0}\rangle - |D^{*-}D^{+}\rangle)\nonumber\\
&&+(|\bar {D}^{0} D^{*0}\rangle - |D^{-} D^{*+}\rangle )],\label{eq43}\\
|{X_{S_{cq}\bar{A}_{cq}/A_{cq}\bar{S}_{cq}}}\rangle&=&\frac{1}{2}[(|{A_{cu}\bar{S}_{cu}}\rangle-|{A_{cd}\bar{S}_{cd}}\rangle)\nonumber\\
&&+(|S_{cu}\bar {A}_{cu}\rangle-|S_{cd}\bar {A}_{cd}\rangle)],\label{eq44}
\end{eqnarray}
recalling that the isospin doublets $(S_{cu},S_{cd})$, $(\bar{S}_{cu},-\bar{S}_{cd})$, $(A_{cu},A_{cd})$, $(\bar{A}_{cu},-\bar{A}_{cd})$, $(D^{(*)0},-D^{(*)+})$ and $(\bar{D}^{(*)0},D^{(*)-})$.

\subsection{the expressions of the effective potentials}

\begin{figure}[htbp]
    \centering
    \includegraphics[width=1.0\linewidth]{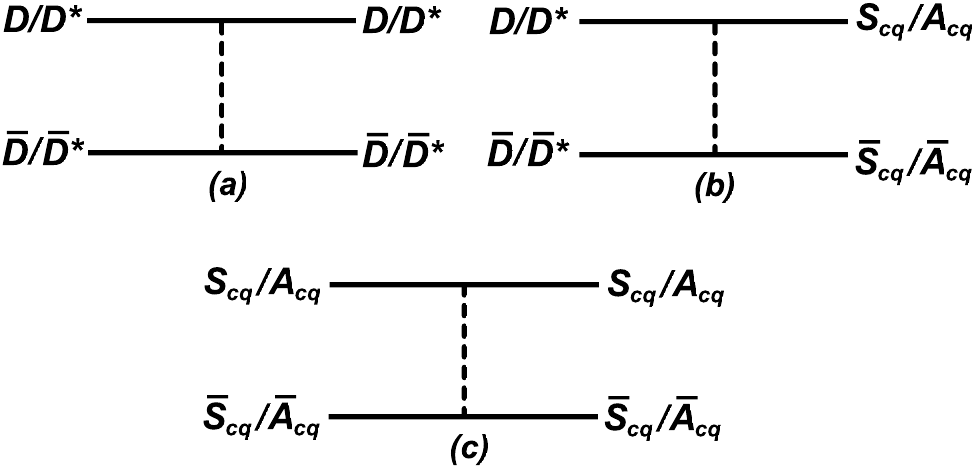}
    \caption{The diagrams needed to be considered in the present work.}
    \label{fig1}
\end{figure} 

With Eqs. \eqref{eq43} and \eqref{eq44}, we derive the total effective potentials are
\begin{eqnarray}
    V_{11}&=&\langle X_{D\bar{D}^*/D^*\bar{D}}|\hat{T}|X_{D\bar{D}^*/D^*\bar{D}}\rangle,\\
    V_{12}&=&V_{21}=\langle X_{S_{cq}\bar{A}_{cq}/A_{cq}\bar{S}_{cq}}|\hat{T}|X_{D\bar{D}^*/D^*\bar{D}}\rangle,\\
    V_{22}&=&\langle X_{S_{cq}\bar{A}_{cq}/A_{cq}\bar{S}_{cq}}|\hat{T}|X_{S_{cq}\bar{A}_{cq}/A_{cq}\bar{S}_{cq}}\rangle.   
\end{eqnarray}
The scattering processes need to be considered are shown in FIG. \ref{fig1}. By using the Feynman rule, we get the subpotentials shown in the following
\begin{eqnarray}
V^{\bar{D}^{*0}D^{0}\rightarrow\bar{D}^{*0}D^{0}}_{\rho^{0}/\omega}&=&-\frac{\beta^{2}g^{2}_{V}}{4m_{p^{*}}m_{p}}m_{D^{0}}m_{D^{*0}}(s-u)\frac{1}{m^{2}_{\rho^{0}/\omega}},\label{eq48}\\
V^{\bar{D}^{*0}D^{0}\rightarrow D^{*-}D^{+}}_{\rho^{+}}&=&-\frac{\beta^{2}g^{2}_{V}}{2m_{p^{*}}m_{p}}\sqrt{ m_{D^{0}}m_{D^{*0}}m_{D^{-}}m_{D^{*-}}}(s-u)\frac{1}{m^{2}_{\rho^{+}}},\nonumber\\\\
V^{D^{*-}D^{+}\rightarrow D^{*-}D^{+}}_{\rho^{0}/\omega}&=&-\frac{\beta^{2}g^{2}_{V}}{4m_{p^{*}}m_{p}}m_{D^{-}}m_{D^{*-}}(s-u)\frac{1}{m^{2}_{\rho^{0}/\omega}},\\
V^{\bar{D}^{0}D^{*0}\rightarrow \bar{D}^{0}D^{*0}}_{\rho^{0}/\omega}&=&-\frac{\beta^{2}g^{2}_{V}}{4m_{p^{*}}m_{p}}m_{D^{0}}m_{D^{*0}}(s-u)\frac{1}{m^{2}_{\rho^{0}/\omega}},\\
V^{\bar{D}^{0}D^{*0}\rightarrow D^{-}D^{*+}}_{\rho^{+}}&=&-\frac{\beta^{2}g^{2}_{V}}{2m_{p^{*}}m_{p}}\sqrt{ m_{D^{0}}m_{D^{*0}}m_{D^{-}}m_{D^{*-}}}(s-u)\frac{1}{m^{2}_{\rho^{+}}},\nonumber\\\\
V^{D^{-}D^{*+}\rightarrow D^{-}D^{*+}}_{\rho^{0}/\omega}&=&-\frac{\beta^{2}g^{2}_{V}}{4m_{p^{*}}m_{p}}m_{D^{-}}m_{D^{*-}}(s-u)\frac{1}{m^{2}_{\rho^{0}/\omega}},\\
V^{\bar{D}^{*0}D^{0}\rightarrow \bar{A}_{cu}S_{cu}}_{A_{uu}}&=&-\sqrt{3} e_{2}\sqrt{m_{D^{0}}m_{D^{*0}}m_{S_{cu}}m_{A_{cu}}}\left[\frac{e_{8}}{2}(s-m_{S_{cu}}^{2}\right.\nonumber\\
& &\left.-m_{A_{cu}}^{2})+\frac{e_{9}}{2}(m_{D^{*0}}^{2}+m_{S_{cu}}^{2}-u)\right]\frac{1}{m_{A_{uu}}^{2}},\label{eq54}\\
V^{\bar{D}^{*0}D^{0}\rightarrow \bar{A}_{cd}S_{cd}}_{A_{ud}}&=&-\frac{\sqrt{3}}{2} e_{2}\sqrt{m_{D^{0}}m_{D^{*0}}m_{S_{cd}}m_{A_{cd}}}\left[\frac{e_{8}}{2}(s-m_{S_{cd}}^{2}\right.\nonumber\\
& &\left.-m_{A_{cd}}^{2})+\frac{e_{9}}{2}(m_{D^{*0}}^{2}+m_{S_{cd}}^{2}-u)\right]\frac{1}{m_{A_{ud}}^{2}},\\
V^{D^{*-}D^{+}\rightarrow \bar{A}_{cu}S_{cu}}_{A_{ud}}&=&-\frac{\sqrt{3}}{2} e_{2}\sqrt{m_{D^{-}}m_{D^{*-}}m_{S_{cu}}m_{A_{cu}}}\left[\frac{e_{8}}{2}(s-m_{S_{cu}}^{2}\right.\nonumber\\
& &\left.-m_{A_{cu}}^{2})+\frac{e_{9}}{2}(m_{D^{*-}}^{2}+m_{S_{cu}}^{2}-u)\right]\frac{1}{m_{A_{ud}}^{2}},\\
V^{D^{*-}D^{+}\rightarrow \bar{A}_{cd}S_{cd}}_{A_{dd}}&=&-\sqrt{3} e_{2}\sqrt{m_{D^{-}}m_{D^{*-}}m_{S_{cd}}m_{A_{cd}}}\left[\frac{e_{8}}{2}(s-m_{S_{cd}}^{2}\right.\nonumber\\
& &\left.-m_{A_{cd}}^{2})+\frac{e_{9}}{2}(m_{D^{*-}}^{2}+m_{S_{cd}}^{2}-u)\right]\frac{1}{m_{A_{dd}}^{2}},\\
V^{\bar{D}^{0}D^{*0}\rightarrow \bar{S}_{cu}A_{cu}}_{A_{uu}}&=&-\sqrt{3} e_{2}\sqrt{m_{D^{0}}m_{D^{*0}}m_{S_{cu}}m_{A_{cu}}}\left[\frac{e_{8}}{2}(s-m_{S_{cu}}^{2}\right.\nonumber\\
& &\left.-m_{A_{cu}}^{2})+\frac{e_{9}}{2}(m_{D^{*0}}^{2}+m_{S_{cu}}^{2}-u)\right]\frac{1}{m_{A_{uu}}^{2}},\\
V^{\bar{D}^{0}D^{*0}\rightarrow \bar{S}_{cd}A_{cd}}_{A_{ud}}&=&-\frac{\sqrt{3}}{2} e_{2}\sqrt{m_{D^{0}}m_{D^{*0}}m_{S_{cd}}m_{A_{cd}}}\left[\frac{e_{8}}{2}(s-m_{S_{cd}}^{2}\right.\nonumber\\& &\left.-m_{A_{cd}}^{2})+\frac{e_{9}}{2}(m_{D^{*0}}^{2}+m_{S_{cd}}^{2}-u)\right]\frac{1}{m_{A_{ud}}^{2}},\\
V^{D^{-}D^{*+}\rightarrow \bar{S}_{cu}A_{cu}}_{A_{ud}}&=&-\frac{\sqrt{3}}{2} e_{2}\sqrt{m_{D^{-}}m_{D^{*-}}m_{S_{cu}}m_{A_{cu}}}\left[\frac{e_{8}}{2}(s-m_{S_{cu}}^{2}\right.\nonumber\\& &\left.-m_{A_{cu}}^{2})+\frac{e_{9}}{2}(m_{D^{*-}}^{2}+m_{S_{cu}}^{2}-u)\right]\frac{1}{m_{A_{ud}}^{2}},\\
V^{D^{-}D^{*+}\rightarrow \bar{S}_{cd}A_{cd}}_{A_{dd}}&=&-\sqrt{3} e_{2}\sqrt{m_{D^{-}}m_{D^{*-}}m_{S_{cd}}m_{A_{cd}}}\left[\frac{e_{8}}{2}(s-m_{S_{cd}}^{2}\right.\nonumber\\& &\left.-m_{A_{cd}}^{2})+\frac{e_{9}}{2}(m_{D^{*-}}^{2}+m_{S_{cd}}^{2}-u)\right]\frac{1}{m_{A_{dd}}^{2}},\label{eq61}
\end{eqnarray}
\begin{eqnarray}
V^{\bar{S}_{cu}A_{cu}\rightarrow \bar{S}_{cu}A_{cu}}_{\rho^{0}/\omega}&=&-h_{1}h_{4}\frac{g_{V}^{2}}{4}m_{A_{cu}}m_{S_{cu}}(s-u)\frac{1}{m_{\rho^{0}/\omega}^{2}},\\ 
V^{\bar{S}_{cu}A_{cu}\rightarrow \bar{S}_{cd}A_{cd}}_{\rho^{+}}&=&-h_{1}h_{4}\frac{g_{V}^{2}}{2}m_{A_{cu}}m_{S_{cu}}(s-u)\frac{1}{m_{\rho^{+}}^{2}},\\
V^{\bar{S}_{cd}A_{cd}\rightarrow \bar{S}_{cd}A_{cd}}_{\rho^{0}/\omega}&=&-h_{1}h_{4}\frac{g_{V}^{2}}{4}m_{A_{cd}}m_{S_{cd}}(s-u)\frac{1}{m_{\rho^{0}/\omega }^{2}},\\
V^{\bar{A}_{cu}S_{cu}\rightarrow \bar{A}_{cu}S_{cu}}_{\rho^{0}/\omega}&=&-h_{1}h_{4}\frac{g_{V}^{2}}{4}m_{S_{cu}}m_{A_{cu}}(s-u)\frac{1}{m_{\rho^{0}/\omega }^{2}},\\
V^{\bar{A}_{cu}S_{cu}\rightarrow \bar{A}_{cd}S_{cd}}_{\rho^{+}}&=&-h_{1}h_{4}\frac{g_{V}^{2}}{2}m_{S_{cu}}m_{A_{cu}}(s-u)\frac{1}{m_{\rho^{+}}^{2}},\\
V^{\bar{A}_{cd}S_{cd}\rightarrow \bar{A}_{cd}S_{cd}}_{\rho^{0}/\omega}&=&-h_{1}h_{4}\frac{g_{V}^{2}}{4}m_{S_{cd}}m_{A_{cd}}(s-u)\frac{1}{m_{\rho^{0}/\omega}^{2}}\label{eq67}
\end{eqnarray}
with the Mandelstam varibles
\begin{eqnarray}
 s&=&(p_{1}+p_{2})^{2},\\
 u&=&(p_{1}-p_{4})^{2}\nonumber\\&=&\frac{m_{1}^{2}+m_{2}^{2}+m_{3}^{2}+m_{4}^{2}}{2}+\frac{(m_{1}^{2}-m_{2}^{2})(m_{3}^{2}-m_{4}^{2})}{2s}-\frac{s}{2}.\nonumber\\
\end{eqnarray}
Note that the subpotentials in Eqs. \eqref{eq48}-\eqref{eq67} have been projected into spin 1. Besides, the factor $\sqrt{3}$ in Eqs. \eqref{eq54}-\eqref{eq61} origin from the color wave function of diquark-antidiquark component.

\section{THE BETHE-SALPETER EQUATION OF THE ON-SHELL FACTORIZED FORM}
 With the effective potentials obtained above, we use the coupled channel Bethe-Salpeter equation of the on-shell factorized form to calculate the $T$-matrix, which is shown below
\begin{eqnarray}
T=(I-VG)^{-1}V.
\end{eqnarray}
Here, $V$ is the matrix of the effective potential , $G$ is the matrix of the two-particle loop function, and the form of the non-zero elements is shown below 
\begin{eqnarray}
G_{ii}=i\int\frac{d^{4}q}{(2\pi)^{4}}\frac{1}{q^{2}-m^{2}_{i1}+i\epsilon}\frac{1}{(P-q)^{2}-m^{2}_{i2}+i\epsilon},
\end{eqnarray}
where $m_{i1}$ and $m_{i2}$ are the masses of the particles in the channel, $i$ represents the label of the channel, $P_{\mu}$ is the four-momentum of the two particles, and $q_{\mu}$ is the four-momentum of one of the particles. The above loop integral is logarithmically divergent, and can be calculated with the three-momentum cutoff regularization. The analytic expression of the cutoff regularization for the loop function has been obtained in the Ref. \cite{Oller:1998hw}, i.e.,
\begin{eqnarray}
	G_{ii}&=&\frac{1}{32\pi^{2}}\left\{\left[\frac{\nu}{s}
	\log\frac{s-\Delta+\nu\sqrt{1+\frac{m_{i1}^{2}}{q_{max}^{2}}}}{-s+\Delta+\nu\sqrt{1+\frac{m_{i1}^{2}}{q_{max}^{2}}}}\right.\right.\nonumber\\\nonumber
	&&\left.+\log\frac{s+\Delta+\nu\sqrt{1+\frac{m_{i2}^{2}}{q_{max}^{2}}}}{-s-\Delta+\nu\sqrt{1+\frac{m_{i2}^{2}}{q_{max}^{2}}}}\right]-\frac{\Delta}{s}\log\frac{m_{i1}^{2}}{m_{i2}^{2}}\\\nonumber
	&&+\frac{2\Delta}{s}\log\frac{1+\sqrt{1+\frac{m_{i1}^{2}}{q_{max}^{2}}}}{1+\sqrt{1+\frac{m_{i2}^{2}}{q_{max}^{2}}}}+\log\frac{m_{i1}^{2}m_{i2}^{2}}{q_{max}^{4}}\\
	&&\left.-2\log\left[\left(1+\sqrt{1+\frac{m_{i1}^{2}}{q_{max}^{2}}}\right)\left(1+\sqrt{1+\frac{m_{i2}^{2}}{q_{max}^{2}}}\right)\right]\right\}.\label{eq69}
\end{eqnarray}
In the above equation, $q_{max}$ stands for the cutoff, $\Delta=m_{i2}^{2}-m_{i1}^{2}$, and $\nu=\sqrt{[s-(m_{i1}+m_{i2})^{2}][s-(m_{i1}-m_{i2})^{2}]}$. Eq. \eqref{eq69} holds on the first Riemann sheet, where bound state can be found, and in order to find resonance poles or virtual states, we need to extrapolate the loop function of a single channel to the second Riemann sheet by a continuation via
\begin{eqnarray}
G_{ii}^{II}=G_{ii}^{I}+i\frac{\nu}{8\pi s}.
\end{eqnarray}
In the present work, two kinds of coupled channels are taken into account, i.e., $D\bar{D}^*/D^*\bar{D}$ and $S_{cq}\bar{A}_{cq}/A_{cq}\bar{S}_{cq}$. For he Bethe-Salpeter equation of the on-shell factorizated form, the Riemann sheet is determined by the matrix of loop functions ($G$ matrix). So we only need to give the definition of Riemann sheets of the $G$ matrix, that is, $G^I=(G^{I}_{11}, G^{I}_{22})$, $G^{II}=(G^{II}_{11}, G^{I}_{22})$, $G^{III}=(G^{II}_{11}, G^{II}_{22})$ and $G^{IV}=(G^{I}_{11}, G^{II}_{22})$. Note that the $G$ matrix is diagonal. 

Besides, since the amplitudes close to a pole behave like
\begin{eqnarray}
T_{ij}=\frac{g_{i}g_{j}}{s-s_{R}},
\end{eqnarray}
we can calculate the effective couplings to each channel.

\section{NUMERICAL RESULTS}
In order to determine the coupling constants in the Lagrangian $\mathcal{L}_1$, we compare it with the Lagrangian in Ref. \cite{Isola:2003fh}, and obtain
\begin{eqnarray}
    a_1=-\frac{\beta}{m_{P}},\ a_2=-2g,\ a_3=-\frac{g}{m_{P^*}},\ a_4=\frac{\beta}{m_{P^*}}.\label{eq40}
\end{eqnarray}
For $g$ in Eq. \eqref{eq40}, we take the full width of $D^{*+}$ from PDG \cite{ParticleDataGroup:2024cfk}, and get $g=0.58\pm 0.01$. The parameter $\beta$ is fixed by the vector meson dominance \cite{Isola:2003fh,Colangelo:1993zq}, i.e., $\beta=0.85$. In the Lagrangians $\mathcal{L}_2$ and $\mathcal{L}_3$, there are two sets of coupling constants $e_i\ (i=1,2,\cdots,9)$ and $h_j\ (j=1,2,\cdots,5)$, which are still unknown. In Ref. \cite{Zhang:2024zbo}, these constants were determined by naively using the quark-pair-creation model with the diquark masses taken from Ref. \cite{Ferretti:2019zyh}. Their values are listed in TABLE \ref{tab2}. Note that we can not fix the signs of $e_3$, $e_5$, $e_6$, $h_2$ and $h_5$ because the relative phase between the amplitudes obtained from the Lagrangians and the quark-pair-creation model can not be determined. For $e_7$, $e_8$ and $e_9$, we get the values of them using the phase in Ref. \cite{Cao:2022rjp} explaining the $Z_{cs}$ states well. We also eliminate the Nambu-Goldstone field $\sigma$ in Eq. \eqref{eq4} by choosing the unitary gauge, i.e., $\sigma=0$ \cite{Harada:2003jx}, such that we do not need to calculate the coupling constant $f_\sigma$. The constant $g_{V}=5.8$ which is determined from the width of $\rho\to\pi\pi$ \cite{Harada:2003jx}.
%The meson decay constant $F_{\pi}=93\ \text{MeV}$, .
\begin{table}
	\caption{The values of the low energy constants in the Lagrangian.}\label{tab2}
	\begin{tabular}{c|c|c|c|c}
		\toprule[1pt]
		 $e_{1}(\text{GeV}^{-1})$&$e_{2}(\text{GeV}^{-1})$&$e_{3}(\text{GeV}^{-2})$& $e_{4}(\text{GeV}^{-1})$ &$e_{5}(\text{GeV}^{-2})$\\ 
		\hline
		 -6.939&4.161&$\pm$1.520&4.039&$\pm$1.588\\
		 \toprule[1pt]
		 $e_{6}(\text{GeV}^{-2})$&$e_{7}(\text{GeV}^{-1})$&$e_{8}(\text{GeV}^{-1})$& $e_{9}(\text{GeV}^{-1})$&  \\
		 \hline
		 $\pm$2.595&2.840&16.778&-11.098&  \\
	    \toprule[1pt]
	    $h_{1}(\text{GeV}^{-1})$&$h_{2}(\text{GeV}^{-2})$&$h_{3}(\text{GeV}^{0})$& $h_{4}(\text{GeV}^{-1})$ &$h_{5}(\text{GeV}^{-1})$\\
	    \hline
	    -0.190& $\pm$0.356&-0.290&-1.632& $\pm$0.032\\
	    \toprule[1pt]
	\end{tabular}
\end{table}

\begin{table}
	\caption{ The pole positions on the second Riemann sheet and the modules of the effective couplings to the $D\bar{D}^{*}$ and $S_{cq}\bar{A}_{cq}$ coupled channels with different cutoffs.}\label{tab3}
	\begin{tabular}{c|c|c|c}
		\toprule[1pt]
		$q_{max}\ (\text{MeV})$&1400&1450&1500\\
		\hline
		Pole (MeV)&3901$\pm$i26&3894$\pm$i22&3888$\pm$i17\\ 
		\hline
		$|g_{D\bar{D}^{*}}|\ (\text{MeV})$& 8606 & 8488 & 8412  \\
		\hline
		$|g_{S_{cq}\bar{A}_{cq}}|\ (\text{MeV})$& 15727 & 15248 & 14874 \\
		\hline
		\toprule[1pt]
	\end{tabular}
\end{table}

As shown in TABLE \ref{tab3}, by solving the Bethe-Salpeter equation, we get the positions of poles and the couplings to each channel depending on different cutoffs for the $D\bar{D}^{*}/S_{cq}\bar{A}_{cq}$ system with the quantum numbers $I^{G}(J^{PC})=1^{+}(1^{+-})$. If the cutoff is chosen from $q_{max}=1400-1500$ MeV, the obtained pole appears on the second Riemann sheet, of which the corresponding mass varies from $3901-3888$ MeV, and the width from $52-34$ MeV. These results are in good agreement with the mass and width of $Z_{c}(3900)$ given in Refs. \cite{BESIII:2020oph,Chen:2023def}, which indicates that the $Z_{c}(3900)$ can be explained as the mixture of $D\bar{D}^{*}$ and $S_{cq}\bar{A}_{cq}$ components. The modules of the couplings $|g_{D\bar{D}^{*}}|$ and $|g_{S_{cq}\bar{A}_{cq}}|$ are calculated as $8.606-8.412$ GeV and $15.727-14.874$ GeV, respectively, which indicates that the $S_{cq}\bar{A}_{cq}$ channel is dominant.

\section{Summary}
In the present work, the interactions of diquarks have been taken into account as well as the mesons interactions within the framework of effective field theory, such that we can introduce the mixing effect of the hadronic molecule and the diquark-antidiquark components in order to reveal the nature of the $Z_{c}(3900)$ state. After we show the method of constructing the Lagrangians describing these interactions, we calculate the effective potentials considering  the $D\bar{D}^{*}/\bar{D}D^{*}$ and $S_{cq}\bar{A}_{cq}/A_{cq}\bar{S}_{cq}$ channels. By solving the Bethe-Salpeter equation with the on-shell parametrized form, we obtain the resonant pole positions on the second Riemann sheet depending on different cutoffs. If the cutoff is chosen from $1400-1500$ MeV, the real part of the obtained pole varies from $3901-3888$ MeV, and the imaginary part from $26-17$ MeV. This result is in good agreement with the mass and width of the $Z_{c}(3900)$ state given in Refs. \cite{BESIII:2020oph,Chen:2023def}, which implies that $Z_c(3900)$ can be explained as the mixture of the $D\bar{D}^{*}/\bar{D}D^{*}$ and $S_{cq}\bar{A}_{cq}/A_{cq}\bar{S}_{cq}$ components. Moreover, the modules of the couplings to the $D\bar{D}^{*}/\bar{D}D^{*}$ and $S_{cq}\bar{A}_{cq}/A_{cq}\bar{S}_{cq}$ channels are obtained as $8.606-8.412$ GeV and $15.727-14.874$ GeV, respectively, which indicates that the diquark-antidiquark component is dominant.

\section*{Acknowledgments}
This project is supported by the National Natural Science Foundation of China (NSFC) under Grants No. 12335001, 11705069 and 11965016, and the National Key Research and Development Program of China under Contract No. 2020YFA0406400.

Yan Ma and De-Shun Zhang contribute equally in this work.
 
\bibliographystyle{apsrev4-1}
\bibliography{ref}

\end{document}